\documentclass[aps,prc,superscriptaddress,showpacs,preprint]{revtex4}

\usepackage{graphicx}

\begin{document}

\title{Dependence of elliptic flow on number of parton degrees of freedom}

\author{Zhe Xu \footnote{xu@th.physik.uni-frankfurt.de}}
\affiliation{Frankfurt Institute for Advanced Studies, 
Ruth-Moufang-Strasse 1, D-60438 Frankfurt am Main, Germany}
\affiliation{Institut f\"{u}r Theoretische Physik, Goethe-Universit\"{a}t
Frankfurt, Max-von-Laue-Strasse 1, D-60438 Frankfurt am
Main, Germany}

\author{Carsten Greiner \footnote{carsten.greiner@th.physik.uni-frankfurt.de}}
\affiliation{Institut f\"{u}r Theoretische Physik, Goethe-Universit\"{a}t
Frankfurt, Max-von-Laue-Strasse 1, D-60438 Frankfurt am
Main, Germany}

\date{\today}

\begin{abstract}
We calculate the elliptic flow parameter $v_2$ for Au+Au collisions at 
$\sqrt{s_{NN}}=200$ GeV employing the parton cascade BAMPS (Boltzmann
Approach of Multiparton Scatterings). Besides gluon interactions including
the bremsstrahlung process, interactions with quarks are considered in
an effective, but approximate way to investigate the dependence of the
collective flow on the number of parton degrees of freedom. We find that
$v_2$ as a function of the transverse momentum $p_T$ is sensitive
to the number of parton degrees of freedom, whereas the $p_T$ averaged
$v_2$ does not. When including quarks, $v_2(p_T)$ shifts to lower $p_T$,
the parton transverse momentum spectra become softer and the mean parton
transverse momenta decrease.
\end{abstract}


\maketitle

\section{Introduction}
\label{intro}
Measurements on collective flow in heavy ion
collisions at the BNL Relativistic Heavy Ion Collider (RHIC) 
\cite{Voloshin:2008dg} showed that the matter produced, the quark
gluon plasma (QGP),  is a nearly perfect 
fluid \cite{Huovinen:2001cy,Schafer:2009dj}. 
Because the strength of the collective flow increases with decreasing 
viscosity, indirect extractions of the shear viscosity have recently been
performed by tuning the shear viscosity (or the QCD coupling) as 
a parameter in viscous hydrodynamic \cite{Romatschke:2007mq,Song:2007ux,Luzum:2008cw,Heinz:2009xj,Teaney:2009qa}
and transport models \cite{Xu:2007jv,Ferini:2008he,Xu:2008av} to 
match the elliptic flow $v_2$. 
A consistent result has been achieved: on a conservative basis
the shear viscosity to the entropy density ratio $\eta/s$ of the QCD matter
at RHIC is less than $0.4$ \cite{Song:2008hj}. Uncertainties stem from
assumptions in modelling the various stages that the matter undergoes
during its evolution.

The results from the transport calculations using the Boltzmann Approach
of Multiparton Scatterings (BAMPS) \cite{Xu:2008av} showed that $v_2(p_T)$ 
as a function of the transverse momentum $p_T$ is lower and the $p_T$ spectra 
are harder than the experimental data, whereas the integral of both
gives a $p_T$ averaged $v_2$, which matches the experimental data. This 
inconsistency may stem from the nature of the rather complicated 
hadronization process
than the simple parton-hadron duality picture used in \cite{Xu:2008av}.
Another reason may lie in the fact that only gluons are considered as 
interacting constituents in the calculations presented in \cite{Xu:2008av}.
Including quark dynamics will increase the number of parton degrees of 
freedom and, thus, may soften the $p_T$ spectra and enhance $v_2(p_T)$. 
In this paper we employ BAMPS and include effective quark degrees of
freedom to investigate how the increase and the equilibration of 
partonic multiplicities affect the elliptic flow and also the $\eta/s$
ratio of the QGP.

This study shall demonstrate whether the buildup of elliptic
flow in ultrarelativistic heavy ion collisions depends on the chemical
equilibration of gluons and quarks. The latter determines the actual
number of constituents in the partonic phase. 
In hydrodynamic calculations it is assumed that the matter stays
in chemical equilibrium and thus the number of constituents is conserved. 
However, the situation in a real heavy ion collision might be considerably
more complicated \cite{Biro:1993qt,Elliott:1999uz}. Quarks are expected
to achieve chemical equilibrium (if this occurs) later than
gluons \cite{Biro:1993qt}.
Even for gluons only, the chemical equilibration will proceed faster
at the collision center than at the region near transverse edge of 
the parton system. The parton chemical equilibration
can well be studied using the microscopic transport model with multiple
scatterings such as BAMPS \cite{Xu:2004mz}, because BAMPS implements
perturbative QCD (pQCD) elastic collisions as well as 
the pQCD based inelastic bremsstrahlung incorporating full detailed balance.

The paper is organized as follows. In Sec. \ref{sec1} we review the parton
cascade BAMPS and introduce effective quarks into BAMPS as a simplification
for real quark dynamics.
In Sec. \ref{sec2} the numerical results are shown to demonstrate the
effect of the parton multiplicities on the elliptic flow of the parton
matter. We summarize in Sec. \ref{sum}.

\section{BAMPS including effective quark degrees of freedom}
\label{sec1}
The detailed model description of the on-shell parton cascade BAMPS can be 
found in Refs. \cite{Xu:2004mz,Xu:2007aa}. In short, the feature of BAMPS is
the successful implementation of particle number changing processes with full
detailed balance. This is ensured by using the stochastic interpretation
of the transition rates in the Boltzmann equations for partons.
Other parton cascade approaches can be found in \cite{Ferini:2008he,Geiger:1991nj,Zhang:1997ej,Molnar:2000jh,Zhou:2006ska,Gombeaud:2007ub}.

The cross section of pQCD gluon elastic scatterings is given 
by \cite{Xu:2004mz,Xu:2008av}
\begin{equation}
\label{cs22}
\frac{d\sigma^{gg\to gg}}{d{\bf q}_{\perp}^2}=
\frac{9 \pi \alpha_s^2}{({\bf q}_{\perp}^2+m_D^2)^2}
\end{equation}
and the effective matrix element of pQCD inspired bremsstrahlung
$gg\leftrightarrow ggg$ is taken in a Gunion-Bertsch form 
\cite{Xu:2004mz,Gunion:1981qs,Biro:1993qt,Wong:1996ta},
\begin{eqnarray}
\label{m23}
| {\cal M}_{gg \to ggg} |^2 &=& \frac{9 g^4}{2}
\frac{s^2}{({\bf q}_{\perp}^2+m_D^2)^2}\,
 \frac{12 g^2 {\bf q}_{\perp}^2}
{{\bf k}_{\perp}^2 [({\bf k}_{\perp}-{\bf q}_{\perp})^2+m_D^2]}\,
\Theta(k_{\perp}\lambda_{\rm mfp}-\cosh y) \,, \\
\label{m32}
| {\cal M}_{ggg \to gg} |^2 &=& | {\cal M}_{gg \to ggg} |^2/d_G\,,
\end{eqnarray}
where $g^2=4\pi\alpha_s$ and $d_G=16$ is the gluon degeneracy factor 
for $N_c=3$. ${\bf q}_{\perp}$ denotes the perpendicular component of 
the momentum transfer, ${\bf k}_{\perp}$ ($k_{\perp}=|{\bf k}_{\perp}|$)
the perpendicular component of the radiated gluon momentum and $y$ its 
rapidity in the center-of-mass frame of the collision, respectively.
The suppression of the bremsstrahlung due to the Landau-Pomeranchuk-Migdal 
effect is effectively taken into account within the Bethe-Heitler regime using 
the step function in Eq. (\ref{m23}). Gluon radiations and absorptions
are only allowed if the formation time of the process, typically 
$\tau=\cosh y/k_{\perp}$, is shorter than the mean free path 
$\lambda_{\rm mfp}$ of the radiated or absorbed gluon. $\lambda_{\rm mfp}$
is calculated self-consistently \cite{Xu:2004mz,Fochler:2008ts},
i.m., $\lambda_{\rm mfp}$ is the inverse of the total collision rate
$R_{gg\to gg}+R_{gg\to ggg}+R_{ggg\to gg}$, where $R_{gg\to ggg}$ and 
$R_{ggg\to gg}$ also depend on $\lambda_{\rm mfp}$ as indicated in
Eqs. (\ref{m23}) and (\ref{m32}) [see Eqs. (\ref{r23}), (\ref{r32}),
(\ref{cs23}), and (\ref{cs32})].

The pQCD interactions are Debye screened. The screening mass is given 
by \cite{Biro:1993qt}
\begin{equation}
\label{md1}
m_D^2({\bf x},t)=\pi \,\alpha_s \, d_G
\int \frac{d^3p}{(2\pi)^3} \ \frac{1}{p} \ 
\left [ N_c f_g({\bf x},t,{\bf p}) + n_f f_q({\bf x},t,{\bf p}) \right ]\,,
\end{equation}
where $n_f$ is the number of quark flavor, and $f_g$ and $f_q$ are
the distribution function of gluons and quarks at a certain quantum state.
For a pure gluon matter $f_q=0$.

The initial parton distribution for BAMPS calculations is the same as 
that chosen in Refs. \cite{Xu:2007jv,Xu:2008av}: an ensemble of gluon
minijets with transverse momenta greater than $p_0=1.4$ GeV, produced
via semihard nucleon-nucleon collisions within a Glauber picture. 
The value of $p_0$ was chosen by matching the parton cascade result
of the final transverse energy per rapidity to the experimental
data \cite{Xu:2007jv,Xu:2008av}. Quark minijets, which take about $20\%$
of the total parton number, are neglected for present studies, in order to
easily make comparisons with the previous results obtained 
in \cite{Xu:2008av}.

To take into account the quark-gluon changing process 
$gg \leftrightarrow q {\bar q}$ and the elastic as well as the 
bremsstrahlung process involving light quarks, we effectively enlarge
the gluon degeneracy factor $d_G$. With a larger $d_G$ than $16$, 
the $ggg \to gg$ process is suppressed [see Eq. (\ref{m32})] and thus more
particles (i.e. quarks) will be produced. This numerical prescription
indicates two assumptions: (1) Quarks are massless. (2) The elastic and 
the bremsstrahlung process involving quarks are identical
with those of gluons according to Eqs. (\ref{cs22})-(\ref{m32}).
These assumptions lead to the same kinetic equilibration of quarks and
gluons, i.e., $f_q \sim f_g$. The second assumption overestimates 
the interaction rate of a quark approximately by the color-charge factor 
$9/4$ due to the dominance of $gg\to gg$, $qg\to qg$,
$gg\leftrightarrow ggg$ and $qg\leftrightarrow qgg$ in kinetic equilibration.
The real equilibration of quarks is expected to be slower than that
of gluons. Moreover, the present prescription used to include quarks cannot
separate quarks from gluons. 
The relative fraction of the quark (or gluon) number to the total parton
number is not known. Quark thermalization by real quark dynamics will be 
investigated in the future \cite{FXG}. The simplified version used here
serves as a tool to study the effect on the elliptic flow when
varying the number of parton degrees of freedom.

The enlarged $d_G$ value can be obtained when considering a fully thermalized
system of quarks and gluons. In this case $f_q=f_g$ and the parton
distribution function is given by $f=d_G f_g+ d_Q f_q=(d_G+d_Q) f_g$,
where $d_Q=24$ is the quark degeneracy factor for $n_f=2$.
To include effective quark degrees of freedom in BAMPS we enlarge $d_G$
from $16$ for a pure gluon system to $40$ for a quark gluon system with
two flavors.

Assuming $f_q=f_g$, which is also valid when the kinetic and chemical
equilibration of quarks and gluons proceeds identically, the screening
mass Eq. (\ref{md1}) becomes
\begin{equation}
\label{md2}
m_D^2({\bf x},t)=\frac{2}{3} \pi \,\alpha_s \, N_c
\int \frac{d^3p}{(2\pi)^3} \ \frac{1}{p} \ f({\bf x},t,{\bf p})\,,
\end{equation}
which is a factor of $2/3$ smaller than the value at the beginning
of the expansion, because initially there are only gluons.
To make a reasonable description of the early stage,
we use instead of Eq. (\ref{md2})
\begin{equation}
\label{md3}
m_D^2({\bf x},t)=\pi \,\alpha_s \, N_c
\int \frac{d^3p}{(2\pi)^3} \ \frac{1}{p} \ f({\bf x},t,{\bf p})
\end{equation}
during the entire parton evolution. Accordingly, $m_D^2$ is overestimated
by a factor of $1.5$ at the late stage of the expansion, when partons
thermalize. The true screening mass will be changing
from Eq. (\ref{md3}) to Eq. (\ref{md2}) in time according to the true
chemical equilibration of gluons and quarks, which, however, cannot be
demonstrated in the present studies. More discussions will be
given in the next section.

As already considered in Refs. \cite{Xu:2007jv,Xu:2008av} for the present
BAMPS calculations, the kinetic freeze-out of particles occurs when the 
local energy density drops below $e_c$, which is assumed to be the critical
value for the occurrence of hadronization. In this paper
we set $e_c=1$ $\rm{GeV\ fm}^{-3}$, which leads to a critical temperature
$T_c=200$ MeV for a pure gluon plasma and $T_c=160$ MeV for a quark gluon
plasma with two quark flavors \cite{text-bose-boltzmann}.
After the freeze-out partons are regarded as massless pions according
to a simple parton-hadron duality picture.
To consider realistic chemical and kinetic
freeze-out a hadronization model and the subsequent hadron cascade should
be included to BAMPS, which will be done in the future.
This, of course, can also have certain influence on the findings in
the next section.

\section{Results}
\label{sec2}
\begin{figure}
\includegraphics[angle=0,width=0.7\textwidth]{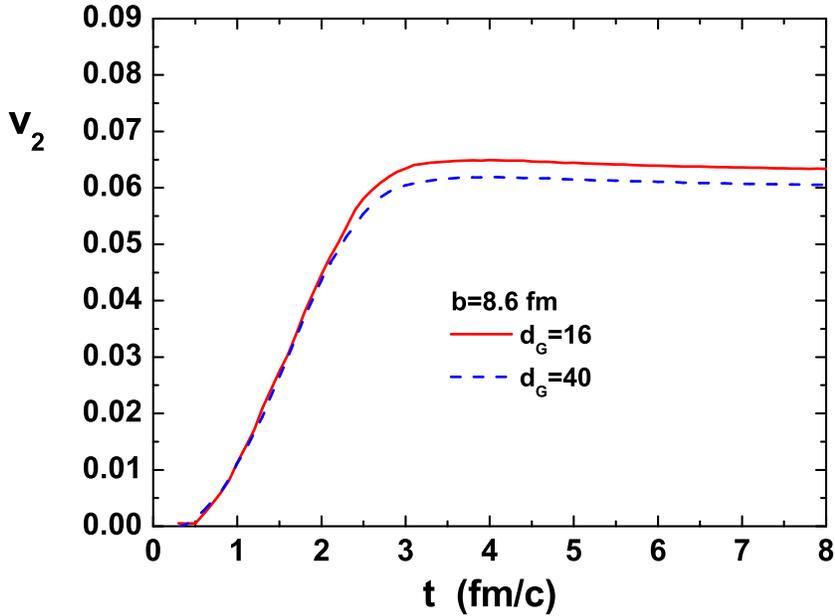}
\caption{(Color online) Time evolution of the $p_T$ averaged $v_2$ 
for particles within $|y| < 1$.}
\label{v2t}
\end{figure}

With the assumptions for the quark dynamics in BAMPS we calculate the space
time evolution of the quark and gluon matter produced in Au+Au collisions at
$\sqrt{s_{NN}}=200$ GeV. The coupling is set to be a constant $\alpha_s=0.6$.
We evaluate the elliptic flow parameter $v_2$
as the average of $(p_x^2-p_y^2)/(p_x^2+p_y^2)$ over particles within
a certain window of momentum rapidity $y=\frac{1}{2} \ln [(E+p_z)/(E-p_z)]$.
Figure \ref{v2t} shows the buildup of the elliptic flow $v_2$
at midrapidity $|y| < 1$ in a Au+Au collision with an
impact parameter of $b=8.6$ fm.
The solid curve gives the result of a pure gluon matter, which was
already obtained from our previous work \cite{Xu:2007jv}. The dashed curve
shows the new result including quarks effectively with the enlarged $d_G=40$.
We see a perfect agreement between the two results except for a difference
of about $5\%$ at later times. The key quantity to understand this
agreement was proposed to be the collision rate per particle
$R=n \langle \sigma \rangle$ (or similarly the Knudsen
number \cite{Bhalerao:2005mm}): For a fixed collision
geometry and a given initial condition the $v_2$ generation depends only
on $R$, but not on the particular details of interactions
among constituent particles. With more degrees of freedom (larger $d_G$)
the total particle density $n$ becomes larger due to the production of quarks.
On the other hand,
the screening mass (\ref{md3}) also becomes larger, which decreases the
total cross section $\langle \sigma \rangle$ of all the interaction
types [see Eqs. (\ref{cs22})-(\ref{m32})]. The increase of $n$ and the
decrease of $\langle \sigma \rangle$ could keep the collision rate
per particle and thus the $v_2$ unchanged. 

To obtain a quantitative answer to this issue, we calculate numerically
the total collision rate per particle
$R_{gg\to gg}+R_{gg\to ggg}+R_{ggg\to gg}$, where
\begin{eqnarray}
\label{r22}
R_{gg\to gg} &=& n \langle v_{\rm rel} \sigma_{gg\to gg} \rangle_2\\
\label{r23}
R_{gg\to ggg} &=& n \langle v_{\rm rel} \sigma_{gg\to ggg} \rangle_2\\
\label{r32}
R_{ggg\to gg} &=& \frac{1}{2} n^2 
\left \langle \frac{I_{ggg\to gg}}{8E_1E_2E_3} \right \rangle_3\,.
\end{eqnarray}
The averages are defined as
\begin{eqnarray}
\langle {\cal O} \rangle_2 &=& \frac{1}{n^2} \int d\Gamma_1 d\Gamma_2 \
f_1(x,p_1) f_2(x,p_2)\ {\cal O} \,,\\ 
\langle {\cal Q} \rangle_3 &=& \frac{1}{n^3} \int d\Gamma_1 d\Gamma_2 
d\Gamma_3 \ f_1(x,p_1) f_2(x,p_2) f_3(x,p_3) \ {\cal Q}
\end{eqnarray}
with
\begin{equation}
\label{ndens}
n= \int d\Gamma_1 \ f(x,p_1) \,,
\end{equation}
where $d\Gamma_j=d^3p_j/(2\pi)^3$, $j=1, 2, 3$, and the local particle 
distribution function $f(x,p)$ is the output from the parton cascade,
$f(x,p)=\sum_i \delta^{(3)}[\vec x -\vec x_i(t)] \ 
\delta^{(3)}[\vec p -\vec p_i]$. The sum runs over all particles
with individual position $\vec x_i$ and momentum $\vec p_i$ at time $t$.

For a $gg\to gg$ or a $gg\to ggg$ process involving two incoming particles
with $(x,p_1)$ and $(x,p_2)$ the relative velocity of the two particles 
is given by $v_{\rm rel}=s/(2E_1E_2)$,
where $s=(p_1+p_2)^2$ is the invariant mass. The total cross sections
read 
\begin{eqnarray}
\label{tcs22}
\sigma_{gg\to gg}&=&\frac{1}{2!}\int_0^{s/4} d{\bf q}_{\perp}^2\,
\frac{d\sigma_{gg\to gg}}{d{\bf q}_{\perp}^2} \,, \\
\label{cs23}
\sigma_{gg \to ggg} &=& \frac{1}{2s} \int
d\Gamma^{'}_1 d\Gamma^{'}_2 d\Gamma^{'}_3
| {\cal M}_{gg \to ggg} |^2 (2\pi)^4
\delta^{(4)}(p_1+p_2-p^{'}_1-p^{'}_2-p^{'}_3)\,,
\end{eqnarray}
where $d\Gamma^{'}_j=d^3p^{'}_j/(2\pi)^3/(2E^{'}_j)$, $j=1, 2, 3$.
For a $ggg\to gg$ process we define
\begin{equation}
\label{cs32}
I_{ggg \to gg} = \frac{1}{2!} \int d\Gamma^{'}_1 d\Gamma^{'}_2
| {\cal M}_{ggg \to gg} |^2 (2\pi)^4
\delta^{(4)}(p_1+p_2+p_3-p^{'}_1-p^{'}_2)\,,
\end{equation}
which is, similar as a cross section, an integral over the final states.
The factor of $1/2!$ in Eqs. (\ref{tcs22}) and (\ref{cs32}) is due to
the fact that the two outgoing gluons are identical particles. 
There is no such a factor ($1/3!$) in $gg\to ggg$ 
processes \cite{Xu:2008av-typos}, because the three outgoing gluons are
distinguished as the radiated gluon ($p^{'}_3$), the radiating gluon 
($p^{'}_2$) and the gluon ($p^{'}_1$), which is only deflected
and does not radiate. Thus, they are kinematically ``distinguishable''
particles.

We note that $n$ in Eq. (\ref{ndens}) is the local particle density
in the center of mass frame of the Au+Au collision (collision frame) 
and only equals the Lorentz invariant particle density if the local 
particle system is at rest, e.g. at the collision center. Also the local
collision rates (\ref{r22})-(\ref{r32}) are calculated in the collision
frame and are smaller than those in the rest frame by a Lorentz factor.

\begin{figure}
\includegraphics[angle=0,width=0.7\textwidth]{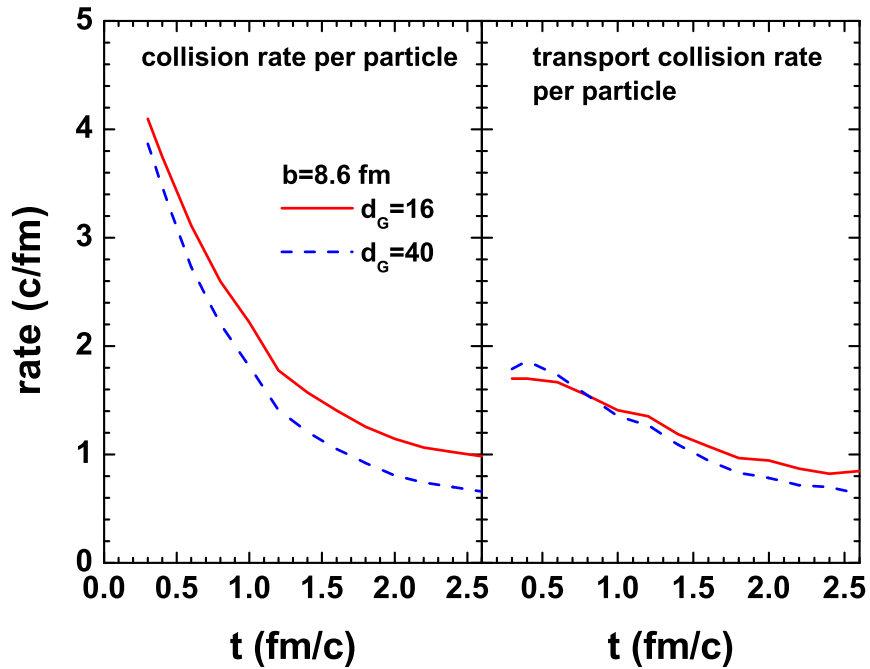}
\caption{(Color online) Time evolution of the collision rate (left) and
the transport collision rate (right) per particle at midrapidity
$|\eta_s| < 0.2$.}
\label{rate}
\end{figure}

The left panel of Fig. \ref{rate} shows the time evolution of the 
total collision rate per particle 
$\langle R_{gg\to gg}+R_{gg\to ggg}+R_{ggg\to gg} \rangle$.
The rate is obtained as the average over all particles in the whole
transverse plan at the central space time rapidity $|\eta_s| < 0.2$. 
$\eta_s$ is defined as $\eta_s=\frac{1}{2} \ln [(t+z)/(t-z)]$. 
The local rate per particle
is calculated according to (\ref{r22})-(\ref{r32}). We have also computed
the average rate by counting interaction events, which occur in BAMPS
within certain time intervals. The results from both calculations agree
with each other.

\begin{figure}
\includegraphics[angle=0,width=0.7\textwidth]{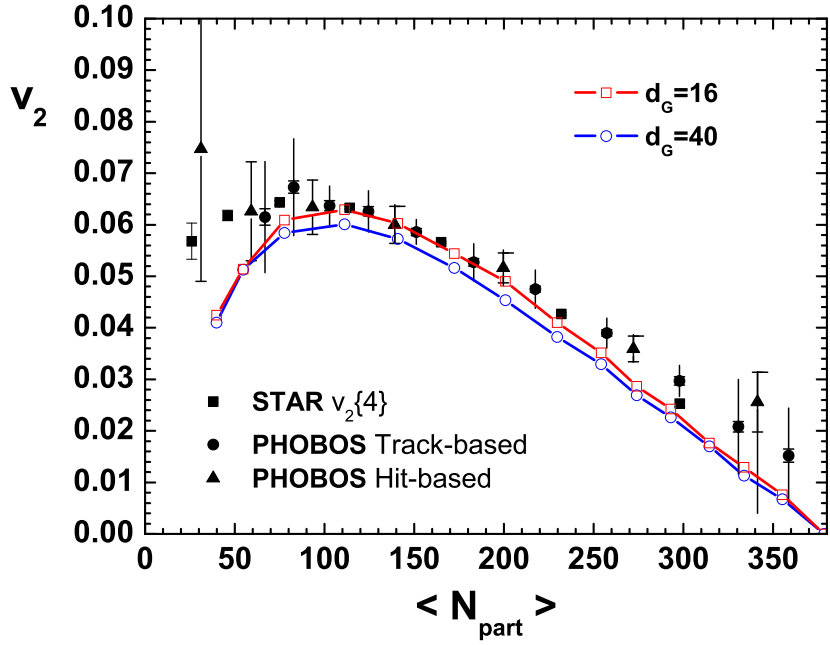}
\caption{(Color online) Elliptic flow parameter $v_2$ ($p_T$ averaged) 
at midrapidity as a function of number of participating nucleons.}
\label{v2total}
\end{figure}

As mentioned before, the collision rate per particle, may determine
the buildup of the elliptic flow $v_2$, which
is shown in Fig. \ref{v2t} within the momentum rapidity $|y| < 1$.
We have also calculated $v_2$ within $|\eta_s| < 0.2$. The agreement of
the two curves seen in Fig. \ref{v2t} is unchanged. The reason why
we show the $v_2$ result within $|y| < 1$ is to compare it with 
the experimental data (see Fig. \ref{v2total}).

From Fig. \ref{rate} we see a moderate difference between the collision
rates in the simulations with $d_G=16$ and $d_G=40$, whereas the average
{\it transport} collision rates per particle shown in the right panel are
much closer to each other. The slightly smaller transport collision
rate in the calculation with $d_G=40$, compared to that with $d_G=16$,
is the reason for the slightly smaller final $v_2$ with $d_G=40$ than
that with $d_G=16$ (see Fig. \ref{v2t}).

The local transport collision rate per particle is given by \cite{Xu:2007aa}
\begin{equation}
\label{trrate}
R_i^{\rm tr}= \frac{\int d\Gamma_1 \frac{p_{z1}^2}{E_1^2} \ C_i[f] -
\langle \frac{p_{z1}^2}{E_1^2} \rangle \int d\Gamma_1  \ C_i[f]}
{n\ \left (\frac{1}{3}- \langle \frac{p_{z1}^2}{E_1^2} \rangle \right )}\,,
\end{equation}
where $\langle {\cal O} \rangle=(1/n) \int d\Gamma_1 \ f_1(x,p_1) {\cal O}$
and $C_i[f]$, $i=gg\to gg, gg\to ggg, ggg\to gg$, denotes the respective
collision term \cite{Xu:2007aa}.
The transport collision rate per particle quantifies the time scale of
thermal equilibration \cite{Xu:2007aa}. Because the angle $\theta_1$ in 
$p_{z1}^2/E_1^2=\cos^2 \theta_1$ enters both the gain and loss term 
of $C_i[f]$, the larger the mean momentum deflection in each collision,
the larger is the transport collision rate and the faster is the momentum
isotropization. The latter leads to a faster buildup of the pressure
gradient and thus to a larger elliptic flow.
Therefore, the transport collision rate, rather than the collision
rate, relates more closely to the buildup of the elliptic flow $v_2$,
as demonstrated in Fig. \ref{rate} in relation with Fig. \ref{v2t}.

The resulting smaller difference in the total transport collision
rates than that in the total collision rates (when varying $d_G$ from
$16$ to $40$) stems from the time evolution of the screening mass $m_D$
[see Eq. (\ref{md3})], which determines not only the cross sections
$\sigma_{gg\to gg}$, $\sigma_{gg\to ggg}$, and $I_{ggg\to gg}$, but also
the collision angle in these processes.
As mentioned after Eq. (\ref{md2}), the choice of $m_D$ is physically
reasonable, because the dynamics at the early stage is dominated by gluons.
On the other hand, $m_D$ is overestimated at the late
times, when ``quarks'' have considerable fraction of the whole parton system.
Thus, the cross sections and the rates at the late times
shown in Fig. \ref{rate} are somewhat smaller than their true values,
when the fully realistic quark dynamics is considered.

In Fig. \ref{v2total} we show the final elliptic flow $v_2$ within $|y| < 1$
for various mean numbers of participating nucleons 
$\langle N_{\rm part} \rangle$ (equivalently impact parameters 
$b$ \cite{Xu:2008av}).
The open squares (or the solid curves) in Fig. \ref{v2total} and all
the following figures depict the results from the calculations with
$d_G=16$, which have already been presented in Ref. \cite{Xu:2008av}.
The open circles show the new results with $d_G=40$, which are slightly
smaller than the values calculated with $d_G=16$ (as also seen in 
Fig. \ref{v2t}), and are still 
comparable with the experimental data at RHIC \cite{Adams:2004bi,Back:2004mh}.

\begin{figure}
\includegraphics[angle=0,width=0.7\textwidth]{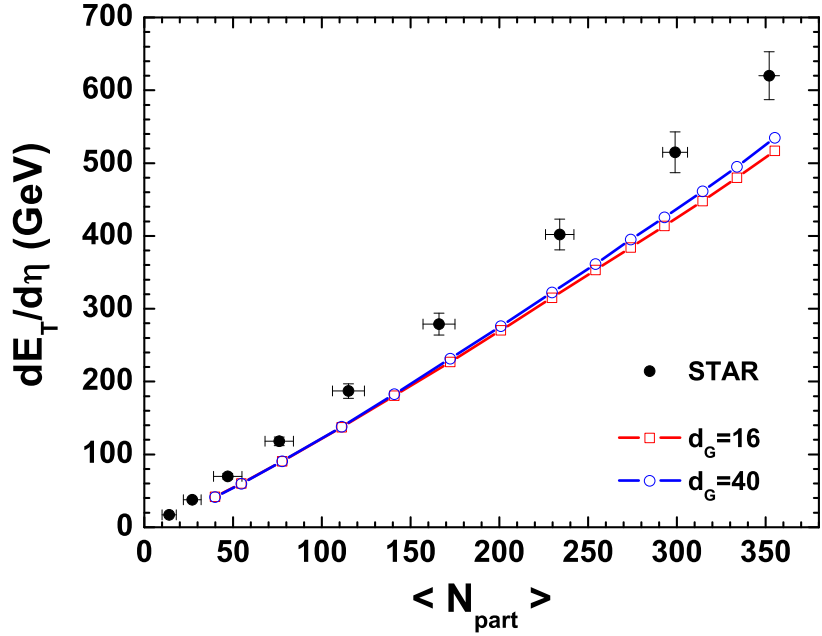}
\caption{(Color online) Transverse energy at midrapidity as a function 
of number of participating nucleons.}
\label{et}
\end{figure}

Figure \ref{et} presents the final transverse energies at midrapidity.
If quarks are included in the initial condition for the parton cascade
calculations, the results would be closer to the experimental 
data \cite{Adams:2004cb}. Comparing the results obtained in the calculations
with $d_G=40$ to those with $d_G=16$, we find a tiny difference. This
implies the same decrease of the transverse energy during the expansion,
which is the consequence of the same viscous effect regardless of the details
of microscopical processes.

\begin{figure}
\includegraphics[angle=0,width=0.7\textwidth]{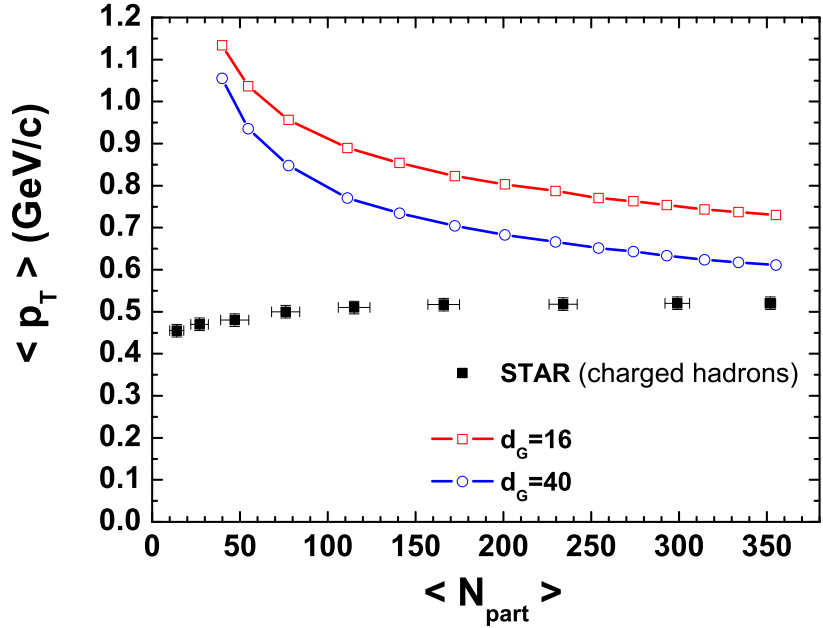}
\caption{(Color online) Mean transverse momentum at midrapidity as a function 
of number of participating nucleons, compared with the STAR 
data \cite{Adams:2004cb}.}
\label{meanpt}
\end{figure}

\begin{figure}
\includegraphics[angle=0,width=0.68\textwidth]{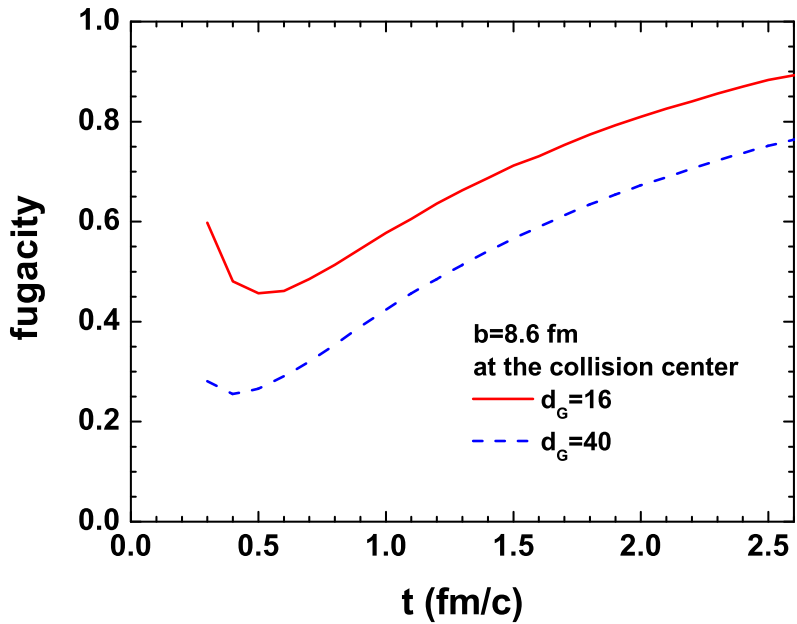}
\caption{(Color online) Time evolution of fugacity.}
\label{fuga}
\end{figure}

On the contrary, the mean parton transverse momentum shown in Fig. \ref{meanpt}
does depend on the details of microscopical processes.
The larger the number of parton degrees of freedom,
the smaller is the final mean transverse momentum. Suppose 
the system is in full thermal equilibrium, we obtain the energy density
$e \sim d_G T^4$ and the particle density 
$n \sim d_G T^3 \sim d_G^{1/4} e^{3/4}$. Similar as the decrease of the
transverse energy the local energy density is not sensitive to $d_G$.
Thus, $n \sim d_G^{1/4}$ and the particle number with $d_G=40$ is
a factor of $1.26$ more than that with $d_G=16$. 
Therefore, to the maximum (in case of thermal equilibrium)
the mean transverse momentum will be reduced by a factor of $1.26$,
when $d_G$ is enlarged from $16$ to $40$. From Fig. \ref{meanpt} we
realize an average reduction by a factor of $1.2$, which is less than 
the maximum reduction. This indicates that full thermalization is not
immediately achieved and particularly the net particle
production (chemical equilibration) in the calculations with $d_G=40$ is
less complete at the freeze-out than for the pure gluon system ($d_G=16$).
The reason is that for the same initial condition the system with $d_G=40$
is initially farther apart from full chemical equilibrium than the system
with $d_G=16$ (see Fig. \ref{fuga}). 
Because the production rates $R_{gg\to ggg}$ are almost the
same in both cases, the chemical equilibration of the system with $d_G=40$
proceeds always behind the process in the system with $d_G=16$.

\begin{figure}
\includegraphics[angle=0,width=0.7\textwidth]{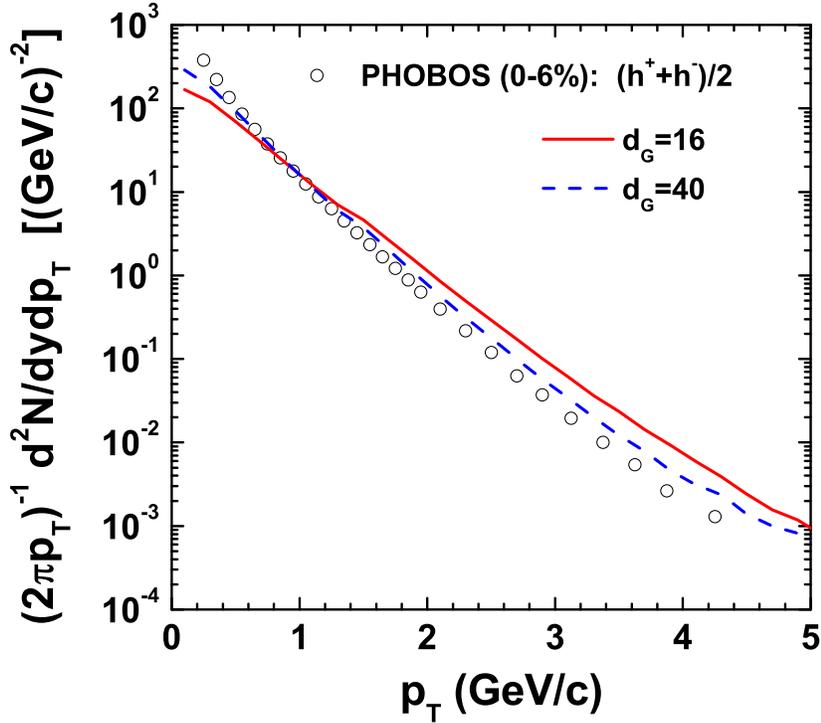}
\caption{(Color online) Momentum spectrum for central collisions.}
\label{dndpt}
\end{figure}

The production of more particles from the same energy content leads to
a smaller temperature at the (chemical) freeze-out. This
clearly relates the softening of the transverse spectrum 
$dN/(p_T dp_T dy)$ shown in Fig. \ref{dndpt} as an example for most 
central collisions.
Both results of the mean transverse momenta and the transverse momentum
spectrum come closer to the experimental data \cite{Adams:2004cb,Back:2003qr},
if the effective quark degrees of freedom are included into the parton
cascade calculations.

\begin{figure}
\includegraphics[angle=0,width=0.7\textwidth]{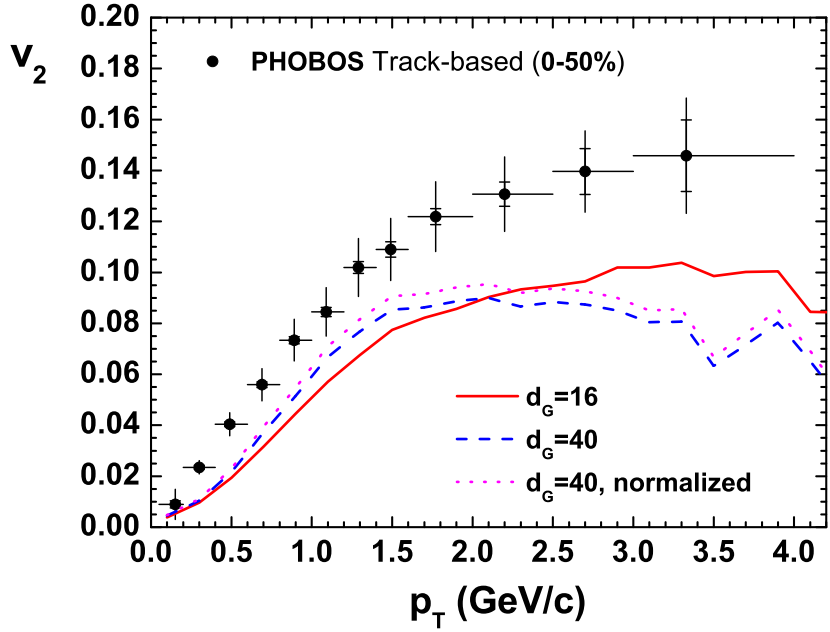}
\caption{(Color online) Momentum dependence of the elliptic flow
$v_2(p_T)$ for the most central $50\%$ collisions.}
\label{v2pt}
\end{figure}

\begin{figure}
\includegraphics[angle=0,width=0.7\textwidth]{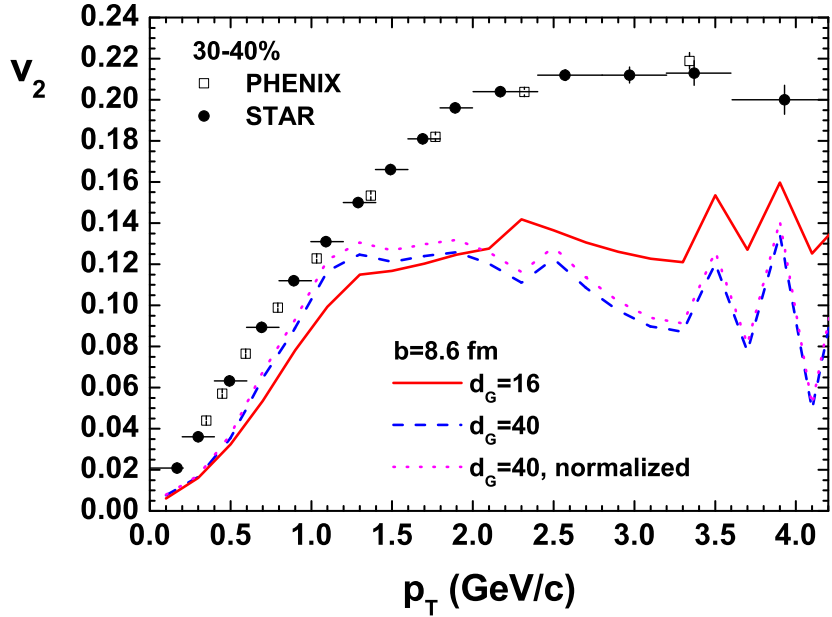}
\caption{(Color online) Momentum dependence of the elliptic flow
$v_2(p_T)$ for the centrality bin $30-40\%$.}
\label{v2pt2}
\end{figure}

On the one hand, the $p_T$ averaged elliptic flow $v_2$ is almost unchanged,
when enlarging $d_G$ from $16$ to $40$. On the other hand, the slope of
the transverse momentum spectrum increases due to the production of more
particles with larger $d_G$. Thus, there must be a simultaneous
change in the differential elliptic flow $v_2(p_T)$, because
$v_2=\int dp_T \ v_2(p_T) \ dN/N/dp_T$. The change of $v_2(p_T)$ is shown
in Fig. \ref{v2pt} for the most central $50\%$ collisions and in 
Fig. \ref{v2pt2} for the mid-central class.
The dotted curves are just the dashed curves times an effective factor,
which enhances the $p_T$ averaged $v_2$ values of the calculations with
$d_G=40$ to be equal with those calculated using $d_G=16$.
The comparison between the dotted and the solid curve is then equivalent
to that of the $v_2(p_T)/v_2$ scaling. Unlike the $p_T$ averaged $v_2$,
$v_2(p_T)$ depends on the number of parton degrees
of freedom: on average, $v_2(p_T)$ moves up $20\%$ or shifts $20\%$ to 
the left toward small momentum and becomes closer to the experimental
data \cite{Adams:2004bi,Back:2004mh,Adare:2006ti}
when quarks are added. This is the major finding of this work.

$v_2(p_T)$ values at large $p_T$ decreases by $20\%$
and is a factor of two smaller than the experimental data. 
Although this has a negligible effect on the $p_T$ averaged
$v_2$, because the parton number there is tiny, as seen in Fig. \ref{dndpt},
it is important to understand hadronization. From our results we
see a bend at $p_T \approx 1.2$ GeV. For higher $p_T$ $v_2(p_T)$ saturates.
If parton recombination models \cite{Lin:2002rw} govern the hadronization
of partons with high $p_T$, $v_2(p_T)$ of mesons will be twice of partons. 
This gives a reasonable explanation of the factor of two difference between
the experimental data and the parton cascade $v_2$ results at high $p_T$.
On the other hand, viscous hydrodynamic calculations \cite{Luzum:2008cw}
showed that $v_2$ at high $p_T$ is very sensitive to the value of shear
viscosity. Also, the difference of the viscous correction to the 
equilibrium phase space distribution between mesons and baryons at 
freeze-out can explain 
constituent quark scaling without quark recombinations \cite{Dusling:2009df}.
Hadronization is still an open issue.

As a final result the shear viscosity $\eta$ from the parton cascade 
simulations is extracted
by using the formula, which has been derived in \cite{Xu:2007ns} and
applied in \cite{Xu:2007jv}:
\begin{equation}
\label{viscos}
\eta \cong \frac{1}{5} n \frac{\langle E(\frac{1}{3}-v_z^2) \rangle}
{\frac{1}{3}-\langle v_z^2 \rangle} \frac{1}{\sum_i R_i^{\rm tr}+
1.5 R_{gg\to ggg}-R_{ggg\to gg}}\,,
\end{equation}
where $i=gg\to gg, gg\to ggg, ggg\to gg$ and $v_z=p_z/E$.
The transport collision rate $R_i^{\rm tr}$ is obtained via
Eq. (\ref{trrate}). The entropy density $s$ is calculated by assuming
the kinetic equilibrium, which gives 
\begin{equation}
\label{entropy}
s=4n-n\,\ln \lambda \,,
\end{equation}
where $\lambda=n/n_{\rm eq}$ with $n_{\rm eq}=d_G T^3/\pi^2$ and
$T=e/(3n)$ defines the gluon fugacity. The true entropy
density is expected to be slightly smaller, because overall kinetic
equilibration cannot be complete in an expanding viscous system.

\begin{figure}
\includegraphics[angle=0,width=0.7\textwidth]{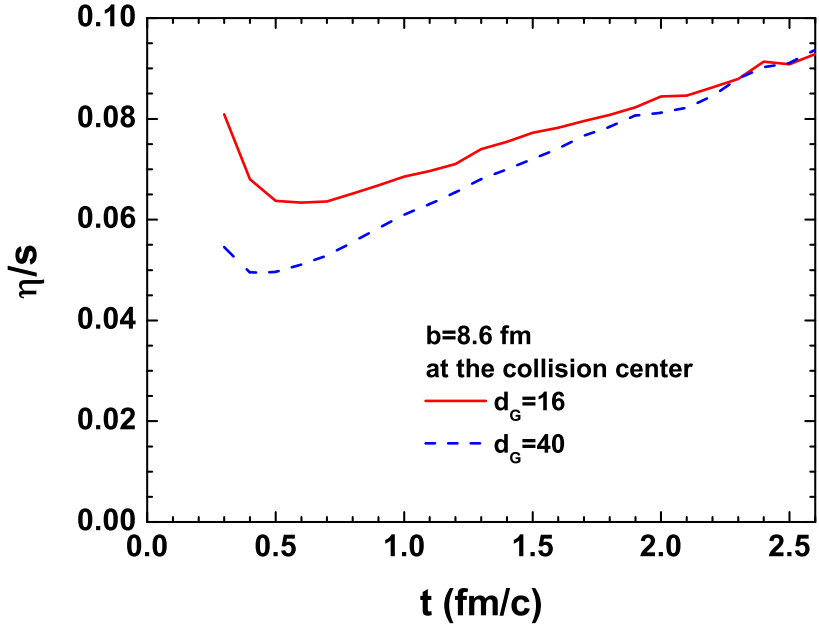}
\caption{(Color online) Time evolution of the ratio of the shear viscosity
to the entropy density.}
\label{etas}
\end{figure}

Figure \ref{etas} shows the $\eta/s$ ratio in the center of a
Au+Au collision at the impact parameter $b=8.6$ fm.
Results from two calculations with $d_G=16$ (solid curve)
and $d_G=40$ (dashed curve) are compared. At early times the $\eta/s$
ratio from the calculation with $d_G=40$ is smaller than that with
$d_G=16$. Because the particle density $n$ in $\eta$ and $s$ cancels, and
quantities $R_i^{\rm tr}$ as shown in Fig. \ref{rate} and 
$\langle E(1/3-v_z^2) \rangle/ (1/3-\langle v_z^2 \rangle)$
are not sensitive to the number of parton degrees of freedom, therefore,
the difference in $\eta/s$ comes mainly from the different chemical 
equilibration in two calculations, i.e., the difference in fugacity
$\lambda$, which appears in Eq. (\ref{viscos}) as 
$1.5R_{gg\to ggg}-R_{ggg\to gg} \approx 1.5 (1-\lambda) R_{gg\to ggg}$
and in Eq. (\ref{entropy}) as $\ln \lambda$.

In Fig. \ref{fuga} we see that the parton fugacity $\lambda(t)$ of the
system with $d_G=40$ is smaller than that with $d_G=16$ due to the increase
of parton degrees of freedom after the initial production. 
Smaller $\lambda$ decreases $\eta$ and increases $s$, and, thus, decreases
the $\eta/s$ ratio, especially at early times
when the system is far from the chemical equilibrium.
We note that the total entropy density of the quark and gluon system
can only be calculated via Eq. (\ref{entropy}), if the $\lambda$ values of
quarks and gluons are the same. In the reality, gluons dominate the
early stage of the plasma and the gluon fugacity should be larger than
the quark's. Therefore, the true entropy density is smaller than that
obtained via Eq. (\ref{entropy}). The true $\eta/s$ ratio of a quark
gluon plasma should be larger than that presented as the dashed curve and 
thus may become closer to the values for a pure gluon plasma.
Moreover, the contribution of the chemical equilibration to $\eta$,
the term $1.5R_{gg\to ggg}-R_{ggg\to gg}$ in Eq. (\ref{viscos}),
does not appear explicitly in the expression derived in Ref. \cite{El:2008yy}
by using the second order Grad's expansion. This may slightly increase $\eta$
compared to the Navier-Stokes value via Eq. (\ref{viscos}), if the system
is out of chemical equilibrium.

\section{Summary}
\label{sum}
Employing the on shell parton cascade BAMPS we have studied the effect
of an increasing number of parton degrees of freedom on the elliptic flow
parameter $v_2$ generated in Au+Au collisions at the RHIC energy 
$\sqrt{s_{NN}}=200$ GeV. 
The initial condition for BAMPS is assumed to be an ensemble of gluon 
minijets. The additional effective quark degrees of freedom during the further 
evolution are created by increasing
the degeneracy factor $d_G$ from $16$ for a pure gluon system to $40$
for a quark gluon system. This prescription indicates the assumption that
quarks and gluons are identical particles. With this assumption for
the BAMPS calculations we have found that the $p_T$ averaged $v_2$ values
and the total transverse energy at midrapidity are
almost unchanged with or without quarks, which is a consequence of the
almost same transport collision rates during the entire expansion.
Second, incorporating quarks the parton multiplicities at freeze-out increase,
which leads to a decrease of the mean parton transverse momenta 
$\langle p_T \rangle$ and a softening of the transverse spectra.
Simultaneously, the differential elliptic flow $v_2(p_T)$ shifts toward lower
momentum. Adding quarks with two flavors brings a $20\%$ effect on
$\langle p_T \rangle$ and $v_2(p_T)$, which is smaller than the maximum
value of $26\%$ due to the incomplete chemical equilibration in the present
studies.
The incomplete chemical equilibration is also the reason for the slightly
smaller $\eta/s$ value in the quark gluon plasma compared with the result for
a pure gluon plasma.

The present prescription for the inclusion of quarks gives an estimate
of interactions with quarks. Bremsstrahlung involving quarks should and will
be explicitly implemented in BAMPS in the future \cite{FXG}, because 
this process is essential for quantifying the thermalization and the 
elliptic flow of quarks.
Moreover, details on the elliptic flow of both quarks and gluons at
the phase transition will provide a more quantitative basis for understanding
the possible effect on the elliptic flow due to the
hadronization of the deconfined matter.

\section*{Acknowledgements}
The BAMPS simulations were performed at the Center for Scientific
Computing of Goethe University.
This work was financially supported by the Helmholtz International Center
for FAIR within the framework of the LOEWE program 
(Landes-Offensive zur
Entwicklung Wissenschaftlich-\"okonomischer Exzellenz)
launched by the State of Hesse.



\begin{thebibliography}{99}
\bibitem{text-bose-boltzmann}
Because Bose enhancement and Pauli blocking are
not implemented in BAMPS, gluons and quarks are assumed to be Boltzmann
particles, i.e., $f_g=f_q={\rm exp}(-E/T)$ in equilibrium.
\end{thebibliography}
\end{document}